\documentclass[%
 reprint,
showpacs,preprintnumbers,
 amsmath,amssymb,
 aps,
]{revtex4-1}

\usepackage{graphicx}
\usepackage{dcolumn}
\usepackage{bm}
\usepackage{hyperref}

\newcommand\dt{$6d\,^2$D$_{3/2}\,$}

\newcommand\ph{$7p\,^2$P$_{1/2}\,$}
\newcommand\sh{$7s\,^2$S$_{1/2}\,$}
\newcommand\szero{$7s\,^1$S$_0\,$}
\newcommand\pone{$7s7p\,^1$P$_1\,$}
\newcommand{\is}{$\delta\nu_{214,M^{'}}$}
\newcommand\trimp{TRI$\mu$P }
\newcommand{\range}{$^{209-214}$}
\newcommand{\raplus}{Ra$^+$}

\begin{document}

\preprint{}

\title{Isotope Shifts of the \dt - \ph~Transition in Trapped Short-Lived \range\raplus}
\author{G. S. Giri}
\email{g.s.giri@rug.nl}
\author{O. O. Versolato}
\author{J. E. van den Berg}
\author{O. B{\"o}ll}
\author{U. Dammalapati}
\author{D. J. van der Hoek}
\author{K. Jungmann}
\author{W. L. Kruithof}
\author{S. M{\"u}ller}
\author{M. Nu\~nez-Portela}
\author{C. J. G. Onderwater}
\author{B. Santra}
\author{R. G. E. Timmermans}
\author{L. W. Wansbeek}
\author{L. Willmann}
\author{H. W. Wilschut}

\affiliation{University of Groningen, Kernfysisch Versneller Instituut, Groningen 9747AA, The Netherlands}

\date{\today}

\begin{abstract}
Laser spectroscopy of short-lived radium isotopes in a linear Paul trap has been performed. The isotope shifts of the \dt - \ph~transition in \range\raplus~were measured, which are sensitive to the short range part of the atomic wavefunctions. The results are essential experimental input for improving the precision of atomic structure calculation. This is indispensable for parity violation in \raplus~aiming at the determination of the weak mixing angle.
\end{abstract}

\maketitle

A measurement of the isotope shifts of the \dt - \ph~transition in a series of trapped, short-lived radium isotopes \range\raplus~with nuclear spins of I = 0, 1/2 and 5/2 is reported. The shifts are complementary to previously measured isotope shifts in this system~\cite{ISOLDE1987isohyper,versolato10,*giricjp2011}. They are sensitive to the short range behavior of the atomic wavefunctions. Of particular importance are the \sh~ground state and the metastable \dt~state in connection with a planned measurement of atomic parity violation (APV) using a single trapped radium ion. The information from optical isotope shift measurements are indispensable to improve atomic structure calculations~\cite{dzubapra01,pal2009,sahoopra07,wansbeekprar08,*versolatocjp2011} and their contributions to APV effects.

The concept for an APV experiment based on a single ion has been worked out for Ba$^+$~\cite{fortsonprl93,Sherman:2008p876,*koerberjpb03}. However, the relative strength of the APV signal is 20 times larger for Ra$^+$~\cite{wansbeekprar08} and an experiment to exploit this enhancement is currently developed within the \trimp research program at KVI. The experiment aims at a 5 fold improved measurement of the weak mixing angle over the sole best APV result in atomic ceasium~\cite{woodsc97,bennettprl99,dereviankoprl09,bouchiatpl82}. 

The radium isotopes $^{212-214}$Ra are produced at the \trimp facility in inverse kinematics by the fusion and evaporation reaction $^{206}$Pb + $^{12}$C $\rightarrow$ $^{(218-x)}$Ra + $x$n, where $x$ is the number of evaporated neutrons~\cite{Shidling2009305}. A $^{204}$Pb beam is used for the production of  $^{209-211}$Ra. The half lives (T$_{1/2}$) of the isotopes are 4.6(2) s for $^{209}$Ra, 3.7(2) s for $^{210}$Ra, 13(2) s for $^{211}$Ra, 13.0(2) s for $^{212}$Ra, 163.8(3.0) s for $^{213}$Ra, and 2.46(3) s for $^{214}$Ra~\cite{bnl-database}. The Ra isotopes are separated from the primary beam and other reaction products in a double magnetic separator~\cite{Berg2006169}. Subsequently they are stopped in a Thermal Ionizer (TI) consisting of a stack of 0.75 $\mu$m thick tungsten foils in a $\sim$ 2500 K hot tungsten cavity \cite{Shidling201011,*Traykov20084478}, and ionized on the hot surfaces. They are electrostatically extracted as a singly charged Ra$^+$ beam with an efficiency $>$8\%~\cite{Shidling2009305}. A Wien filter in the ion transport system suppresses other ion species from the TI. The \raplus~beam is injected into a gas-filled Radio Frequency Quadrupole (RFQ) cooler. The segmented linear quadrupole with an electrode tip spacing of 5 mm is operated as a trap by applying suitable potentials to the segments along the axis of the RFQ. An RF amplitude of 190 V at 500 kHz between the opposite electrodes yields an effective potential depth of 13 V. An axial potential depth of 10 V is chosen for optimal trapping efficiency. The ions are thermalized by Ne buffer gas  with a pressure of 1-5 $\times$ 10$^{-2}$ mbar. The typical number of ions in the trap is about 1000 for $^{212}$\raplus.

\begin{figure}[h]
\includegraphics[width = 8.6 cm, angle = 0]{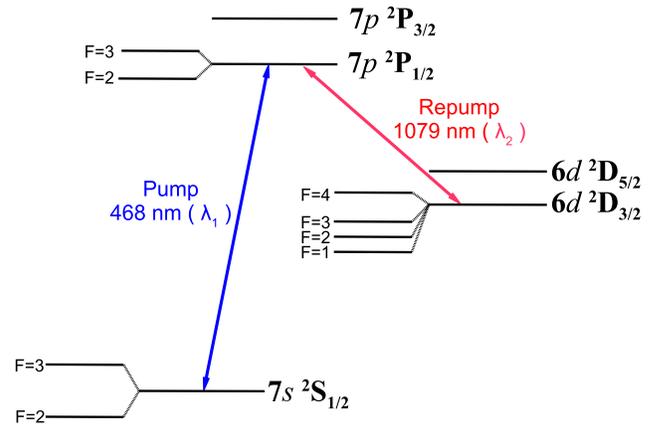}
\caption{(Color online) Level scheme of $^{209,211}$Ra$^+$ with nuclear spin I=5/2. The hyperfine structure splitting is enlarged for visibility.} \label{level1}
\end{figure}

\begin{figure}[]
\includegraphics[width = 8.6 cm, angle = 0]{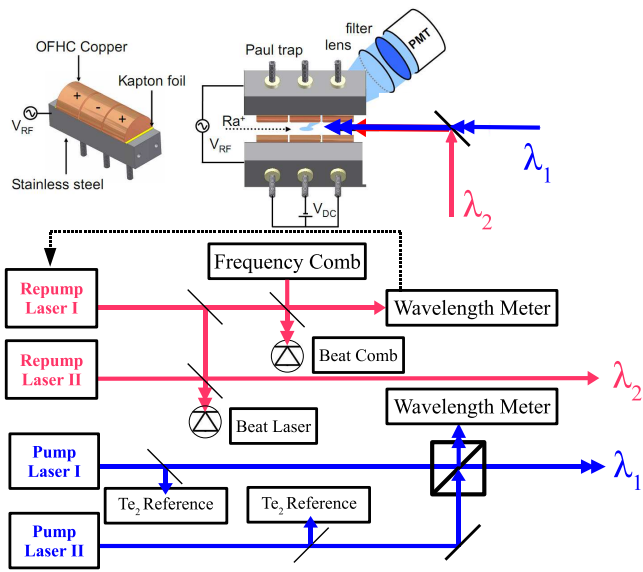}
\caption{(Color online) Schematics for ion trapping and laser spectroscopy. Repump laser I (auxiliary laser) is continuously locked to the wavelength meter. This is also cross-referenced to the frequency comb and serves as a reference laser. Repump laser II is used for spectroscopy. Both the pump lasers are used for spectroscopy and they are monitored with a tellurium (Te$_2$) molecular reference.} 
\label{setup}
\end{figure}

Laser spectroscopy is performed by driving optical transitions at wavelengths $\lambda_1$ and $\lambda_2$ (Fig.~\ref{level1}). Light from two extended cavity diode lasers at wavelength $\lambda_1$ is overlapped and transported to the ion trap in a single mode optical fiber. A separate single mode optical fiber transports  light from an extended cavity diode laser at wavelength $\lambda_2$ to the trap. The laser beams at different wavelength are overlapped on a dichroic mirror and aligned along the axis of the trap in order to minimize stray light. Typical laser intensities at the trap center are 200 $\mu$W/mm$^2$($\lambda_1$) and 600 $\mu$W/mm$^2$ ($\lambda_2$). The transitions are detected by fluorescence at wavelength $\lambda_1$ by an imaging lens system and a photo multiplier tube. A low-pass filter with 80\% transmission for wavelengths shorter than 500 nm suppresses unwanted stray light. The setup is shown in Fig.~\ref{setup}.

The wavelengths of the lasers are monitored by two commercial wavelength meters with an absolute frequency uncertainty of 50 MHz. This was achieved by calibrating them against the molecular iodine I$_2$ reference line R116(2-9)a15 at $\mathrm{\nu_{I_2}}$ = 419,686,834(3) MHz employing saturated absorption spectroscopy~\cite{dammalapati2009}. The laser at wavelength $\lambda_1$ is stabilized to a fraction of the Doppler broadened linewidth of about 300 MHz (FWHM)  while the frequency of the laser at $\lambda_2$ is determined with the help of an optical frequency comb. The latter is achieved by employing an auxiliary laser at wavelength $\lambda_2$, which is referenced against the frequency comb by a beatnote $f_\mathrm{BeatComb}$ = $\nu_\mathrm{aux} - \nu_{m^\mathrm{th}}$, where $\nu_\mathrm{aux}$ is the frequency of auxiliary laser and $\nu_{m^\mathrm{th}}$ is the frequency of the $m^\mathrm{th}$ comb line. The signal-to-noise ratio of the beatnote is typically 40 dB with 1 MHz bandwidth. The frequency $\nu_\mathrm{Laser}$ of the laser which drives the \dt - \ph~transition is determined by the beatnote frequency $f_\mathrm{BeatLaser}$ = $\nu_\mathrm{Laser} - \nu_\mathrm{aux}$ with the auxiliary laser, yielding
\begin{equation}
\nu_\mathrm{Laser}=m\times f_\mathrm{REP} + f_\mathrm{CEO} + f_\mathrm{BeatComb} + f_\mathrm{BeatLaser}.
\end{equation}
Here, $m$ is the mode number, $f_\mathrm{REP}$ = 250,041,000 Hz is the repetition rate and $f_\mathrm{CEO}$=20,000,000 Hz is the carrier-envelope-offset frequency of the comb. The accuracy of the frequency comb is derived from a GPS stabilized Rb clock to be better than 10$^{-11}$ which is not limiting the precision of the results presented in this paper. The sign of $f_\mathrm{BeatLaser}$ is either positive or negative depending on the detuning of the scanning laser with respect to the auxiliary laser. The course determination of the frequency with the calibrated wavelength meter yields a mode number $m$ = 1111032. The uncertainty is only a fraction (0.2) of one mode spacing due to the calibration accuracy of the wavelength meter. This results in the frequency of the $m^{\mathrm{th}}$ comb line $\nu_{m^{\mathrm{th}}}$ = 277,803,572.31(3) MHz, where the uncertainty arises from the Rb clock. 

\begin{table}[]
\caption{Measured frequencies of the transitions in even isotopes with respect to the frequency $\nu_{m^{\mathrm{th}}}$ of the  $m^\mathrm{th}$ comb line in MHz.}
\begin{ruledtabular}
\begin{tabular}{llll}
	& $^{210}$\raplus& $^{212}$\raplus& $^{214}$\raplus	\\
\colrule
\dt - \ph		& 200(11)& 1059(2)& 2084(11)	\\
\end{tabular}
\end{ruledtabular}
\label{peak-positions-even}
\end{table}
\begin{table}[]
\caption{Measured frequencies of the transitions in odd isotopes with respect to the frequency $\nu_{m^{\mathrm{th}}}$ of the $m^\mathrm{th}$ comb line in MHz.}
\begin{ruledtabular}
\begin{tabular}{llll}
\dt - \ph		& $^{209}$\raplus& $^{211}$\raplus& $^{213}$\raplus	\\
\colrule
F=4$\rightarrow$F=3		& 455(30)		& 1363(3)	& 	-		\\
F=3$\rightarrow$F=3		& 1130(58)		& 2049(10)	& 	-		\\
F=2$\rightarrow$F=3		& 1527(68)		& 2456(12)	& 	-		\\
F=1$\rightarrow$F=0		& 	-	& -	& $-$1356(2)	\\
\end{tabular}
\end{ruledtabular}
\label{peak-positions-odd}
\end{table}

\begin{figure}[]
\centering
\includegraphics[width = 8.6 cm, angle=0]{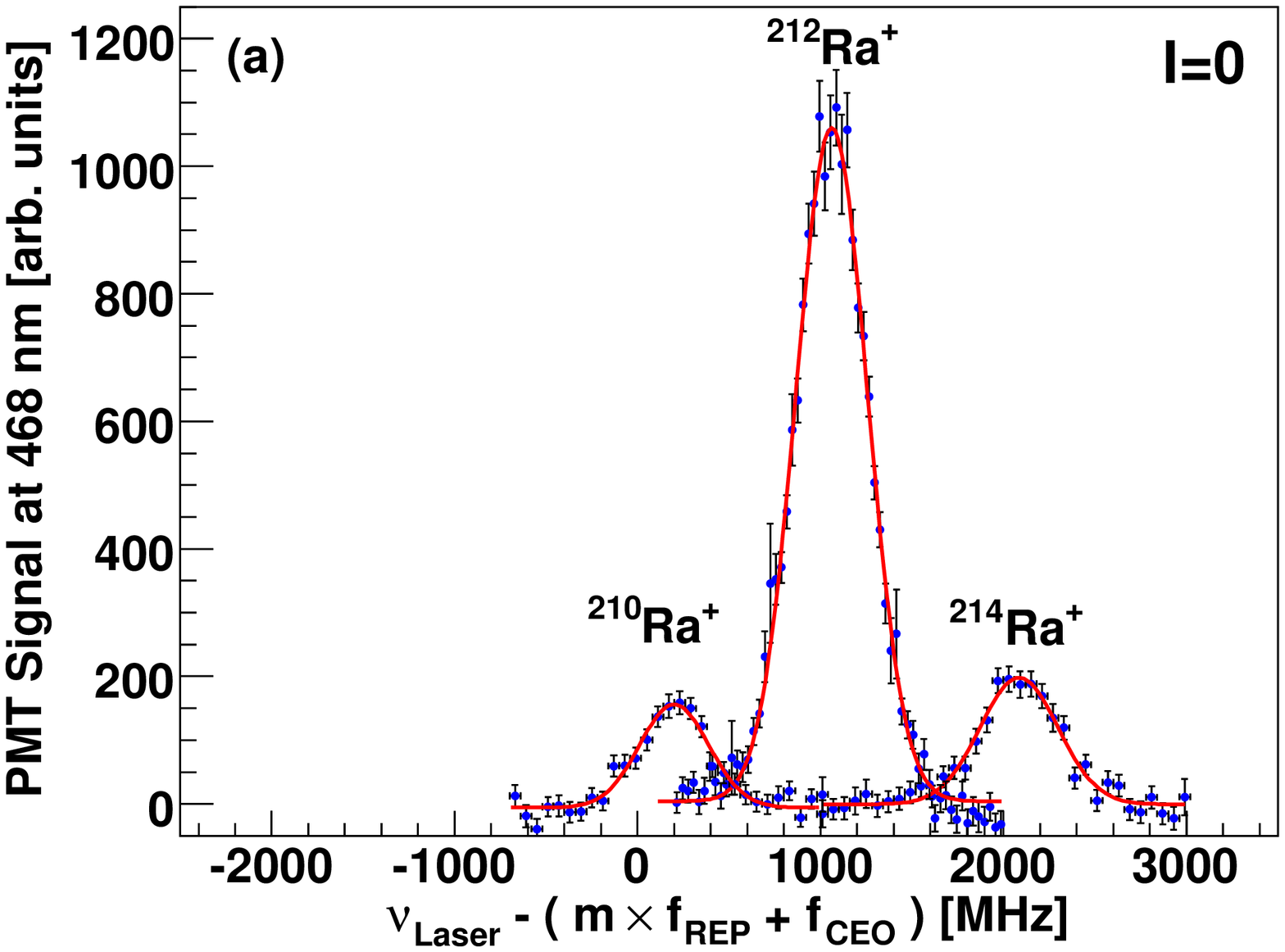}
\includegraphics[width = 8.6 cm, angle=0]{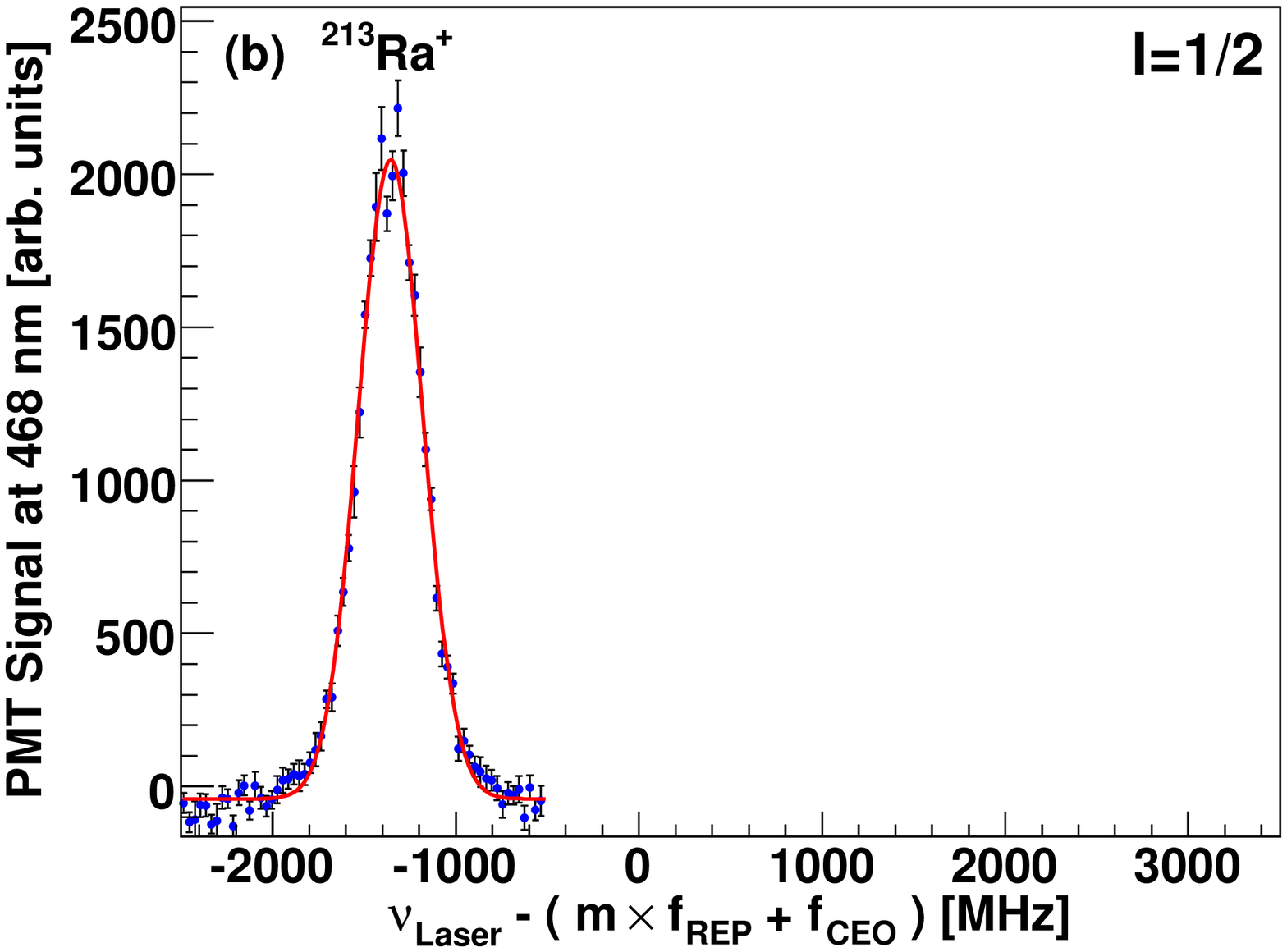}
\includegraphics[width = 8.6 cm, angle=0]{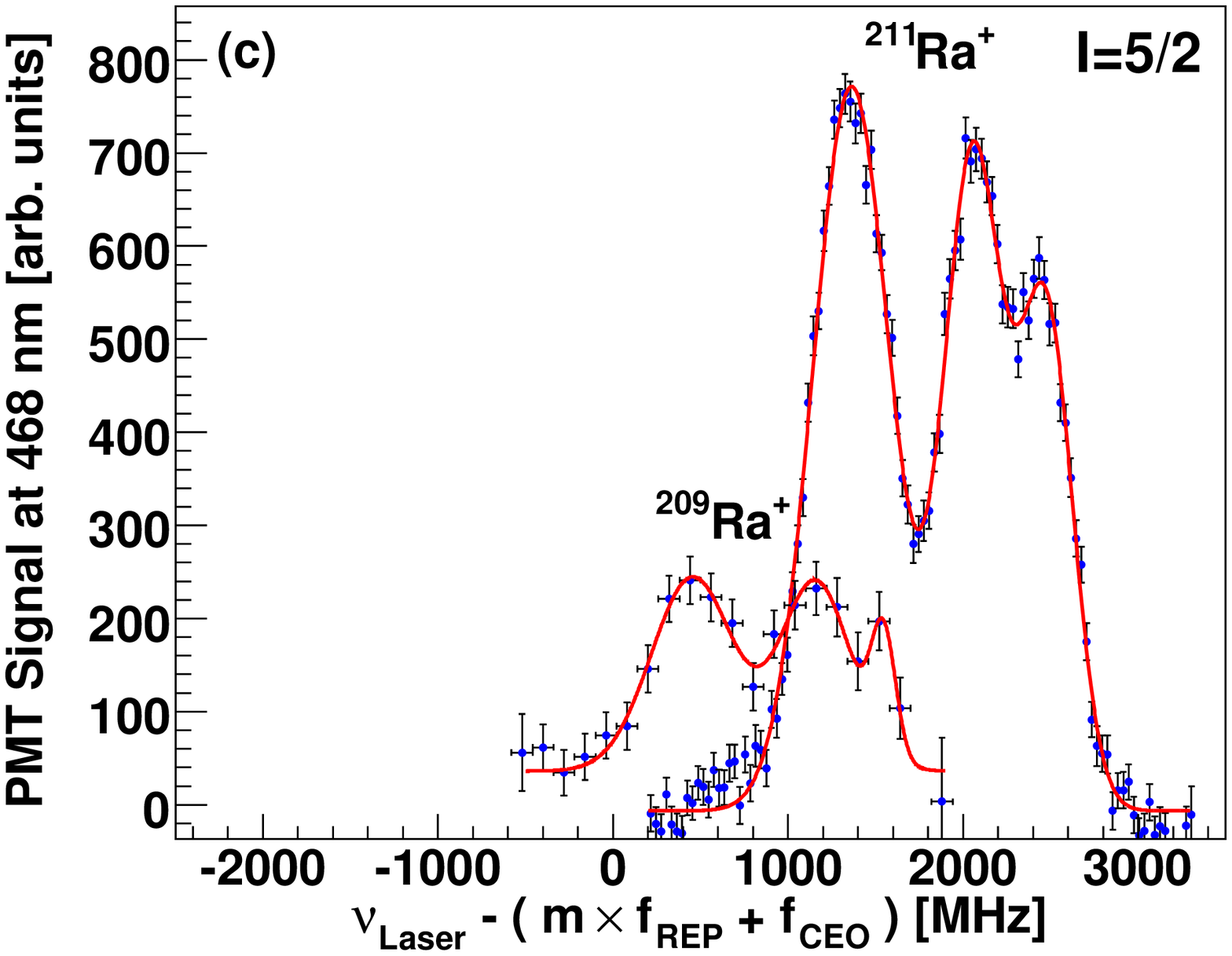}
\caption{(Color online) \dt - \ph~transitions in $^{209-214}$\raplus. The horizontal axis shows the laser frequency offset from the frequency of 1111032$^{\mathrm{th}}$ comb line. The solid lines represent a fit with one or three Gaussian functions. (a) Isotopes with I=0. (b) Isotope with I=1/2: The \dt(F=1) - \ph(F=0) transition is shown. (c) Isotopes with I=5/2: The \dt(F=4,3,2) - \ph(F=3) transitions are shown, where all peaks at low frequency correspond to the \dt(F=4) - \ph(F=3) transitions.}
\label{209-214}
\end{figure}

In the experiment it is sufficient to drive the transitions by one laser at the pump wavelength $\lambda_1$ and one at the repump wavelength $\lambda_2$ for isotopes with I=0. For isotopes with I$\neq$0, both hyperfine ground states are pumped simultaneously by individual lasers, since the hyperfine splitting of \sh~state of 15-25 GHz is much larger than the observed linewidth. The \sh - \ph~transition is continuously driven. The excited \ph~state decays into \dt~state with branching ratio of about 1 in 10. The \dt~state decays into the ground state by radiative decay and collision induced transitions which depend on the type of buffer gas and its pressure~\cite{hermanni:1989}. This is much slower than the resonant laser driven transfer to the ground state. Furthermore, the buffer gas provides mixing of the hyperfine states, which prevents optical pumping to a dark state. The repump transition is driven by a single laser, since the hyperfine splitting of \dt~state is less than a GHz. To obtain a good signal-to-noise ratio, many scans over the resonances are averaged. A set of \dt~hyperfine states can be populated by selecting \sh - \ph~hyperfine transitions. The \dt(F=1) state in $^{213}$\raplus (I=1/2) is populated by choosing the \sh(F=1) - \ph(F=0) transition, thereby forming a closed cycle. The \dt(F=2,3,4) states in $^{209,211}$\raplus (I=5/2) are populated by selectively pumping via the \ph(F=3) state. The observed transitions have been calibrated to a common frequency axis (Fig.~\ref{209-214}a to \ref{209-214}c). The position of each transition is determined by fitting the data with Gaussian profiles. The measured transitions frequencies are listed in Table~\ref{peak-positions-even} and~\ref{peak-positions-odd}. The quoted uncertainties are limited by the statistical precision of the data. The uncertainties due to possible systematic effects such as residual Doppler shifts, pressures shifts, lifetimes, and laser linewidths have been estimated to be on the level of 1 MHz, which is very small compared to the statistical uncertainties.

The isotope shift of a spectral line is commonly parameterized as~\cite{pendril1992}
\begin{equation}
\delta\nu_{MM^{'}}=(K_{\mathrm{NMS}}+K_{\mathrm{SMS}})\frac{M-M^{'}}{MM^{'}}+F_ {\mathrm{FS}}~\delta\langle r^2\rangle_{MM^{'}}, \label{isotopeshift}
\end{equation}
where $\delta\nu_{MM^{'}}=\nu_M-\nu_{M^{'}}$. M and M$^{'}$ are the masses of the reference isotope and the isotope of interest respectively, both in atomic mass units. $K_{\mathrm{NMS}}$ and $K_{\mathrm{SMS}}$ are the normal and specific mass shift components and $F_ {\mathrm{FS}}$ is the field shift component. $\delta\langle r^2\rangle_{MM^{'}}$ is the difference in mean square nuclear charge radii, which is common to all transitions for a pair of nuclei. Both the specific mass shift and the field shift are characteristic to a particular transition. Here they are defined as the shift of the lower level, minus the shift of the upper level. The normal mass shift can be calculated from the expression $K_ {\mathrm{NMS}} = \nu m_e$, where $\nu$ is the transition frequency and $m_e$ is the mass of electron in atomic mass units.

A King plot analysis~\cite{king1963} is used to separate the specific mass shift and the field shift components. Here the transformed isotope shifts ($\Delta\nu_{MM^{'}}^{\mathrm{King}}$) are used, which are obtained by subtracting the normal mass shift component from the experimentally measured isotope shifts and multiplying both sides of Eq.~(\ref{isotopeshift}) by $MM^{'}/(M-M^{'})$. Eq.~(\ref{isotopeshift}) is rewritten as
\begin{subequations}
\begin{align}
\Delta\nu_{MM^{'}}^{\mathrm{King}}&=K_{\mathrm{SMS}} + F_ {\mathrm{FS}}~\delta\langle r^2\rangle_{MM^{'}}\frac{MM^{'}}{M-M^{'}}\\
&= \delta\nu_{MM^{'}}\frac{MM^{'}}{M-M^{'}}-K_{\mathrm{NMS}}. \label{transformed-is} 
\end{align}
\end{subequations}
In a comparison between two different optical transitions $i$ and $j$, the transformed isotope shifts can be written as
\begin{eqnarray}
\Delta\nu_{MM^{'}}^{j, \mathrm{King}} = \frac{F_{FS}^j}{F_{FS}^i} \Delta\nu_{MM^{'}}^{i, \mathrm{King}}+ K_{SMS}^j-\frac{F_{FS}^j}{F_{FS}^i}K_{SMS}^i.
\label{transformed-is-ratio}
\end{eqnarray}
This is essentially a linear relation of the transformed isotope shifts of one transition ($\Delta\nu_{MM^{'}}^{j, \mathrm{King}}$) against the corresponding shifts of the other transition ($\Delta\nu_{MM^{'}}^{i, \mathrm{King}}$). The slope yields the ratio of field shifts and the difference in specific mass shifts appears as the crossing with the abscissa. 

$^{214}$\raplus~has been chosen as the reference isotope in order to be consistent with previous work~\cite{ISOLDE1987isohyper}. The difference in transition frequencies for the even isotopes yields directly the isotope shift. For $^{209, 211, 213}$\raplus (I$\neq$0) the isotope shifts are given between the centers of gravity of the \dt~and \ph~states. This requires the magnetic dipole $A$ and the electric quadrupole $B$ hyperfine constants for those states (Table~\ref{isotope-shift}). For the \dt~state, the hyperfine constants have been derived from measured hyperfine intervals~\cite{versolato10,versolato2011}. The hyperfine constants $A$ for the \ph~state are taken from~\cite{ISOLDE1987isohyper}. For the case of $^{209}$\raplus~no value has been reported. It is derived with the nuclear magnetic moments $\mu$ ~\cite{ISOLDEmomentsw} using
\begin{eqnarray}
\frac{A(7p\,^2P_{1/2},^{209}\mathrm{Ra^+})}{A(7p\,^2P_{1/2},^{213}\mathrm{Ra^+})}=\frac{I(^{213}\mathrm{Ra^+})}{I(^{209}\mathrm{Ra^+})} \times \frac{\mu(^{209}\mathrm{Ra^+})}{\mu(^{213}\mathrm{Ra^+})}.
\label{ratio-method}
\end{eqnarray}
The measured isotope shifts for all the even and odd isotopes (\range\raplus) and the relevant hyperfine constants for the \dt~and \ph~states in case of isotopes with I$\neq$0 are given in Table~\ref{isotope-shift}.

\begin{table}[]
\caption{Isotope shifts of \dt - \ph~transition in \range\raplus~with the relevant hyperfine constants. The comparatively large error bar on the isotope shift of $^{209}$\raplus~is a consequence of low signal-to-noise ratio due to lower yield and shorter life time of the isotope.}
\begin{ruledtabular}
\begin{tabular}{lllll}
Mass	&	A(\dt)	&	B(\dt)	& 	A(\ph)	& \is 		\\
Number	&	MHz	&	MHz	& 	MHz	& MHz		\\ 
\colrule
214 	&	-	&	-	& 	-	&  0	\\
213	&	528(5)~\cite{versolato10}	& 	-	&	4525(5)~\cite{ISOLDE1987isohyper}	& 707(14)	\\
212 	&	-	&	-	&	-	& 1025(12)	\\
211	&	151(2)~\cite{versolato2011}  &	103(6)~\cite{versolato2011}	&  1299.7(0.8)~\cite{ISOLDE1987isohyper}	& 1755(14)	\\
210	&	- 	&	-	&	-	&  1884(16)	\\
209	&	148(10)~\cite{versolato2011}	&	104(38)~\cite{versolato2011}	&  1276(20)\footnote{Calculated using Eq.~\ref{ratio-method}}	& 2645(56) \\	
\end{tabular}
\end{ruledtabular}
\label{isotope-shift}
\end{table}

\begin{figure}[]
\includegraphics[width = 8.6 cm, angle=0]{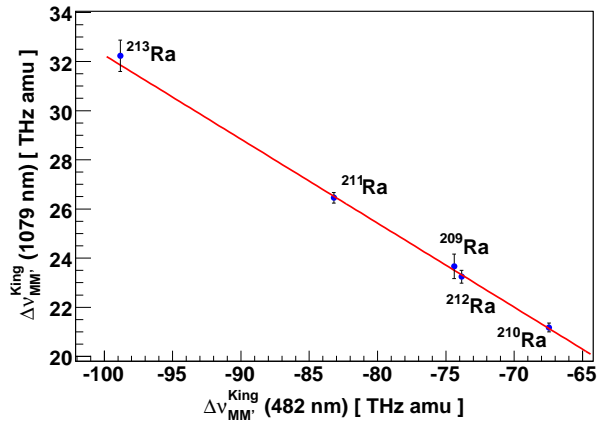}
\caption{(Color online) King plot of the transformed isotope shifts (Eq.~\ref{transformed-is} and~\ref{transformed-is-ratio}) of measured \dt - \ph~transition (1079 nm) for the isotope chain \range\raplus~against the corresponding shifts of the \szero - \pone~transition (482 nm) in neutral radium isotopes. The uncertainties of the measured shifts of the 482 nm line in neutral radium are below 8 MHz. This leads to error bars which are smaller than the size of markers in this plot.}
\label{kingplot}
\end{figure}

A King plot of the transformed isotope shifts of the measured \dt - \ph~transition against the corresponding shifts of the \szero - \pone~transition in neutral radium~\cite{ISOLDE1987isohyper} is shown in Fig.~\ref{kingplot}. The values for the atomic masses of the isotopes are taken from~\cite{nist-database}. The normal mass shift coefficient of the \dt - \ph~transition in ionic radium is $K_ {\mathrm{NMS}}$ = 152.4 GHz amu. For the \szero - \pone~transition in neutral radium, the normal mass shift coefficient is $K_ {\mathrm{NMS}}$ = 340.8 GHz amu. Plotting the transformed isotope shifts against each other show that the data satisfies a linear relation within the measurement uncertainties (Fig.~\ref{kingplot}). The slope determines the ratio of field shift coefficients, $F_{FS}^{1079~\mathrm{nm}}/F_{FS}^{482~\mathrm{nm}}$ = $-$0.342(15). The abscissa determines the difference of the specific mass shift to be $-$1.9(1.1) THz amu.

In summary, online laser spectroscopy has been performed using a range of trapped, short-lived radium isotopes \range\raplus. Isotope shifts of \dt - \ph~transition have been determined. A comparison between these isotope shifts with another optical transition is made using a King plot analysis. This yields the ratio of the field shifts and difference in the specific mass shifts between the two transitions. This provides a test of the atomic theory of \raplus~at the few percent level, in particular of the \dt~states which are relevant for an APV measurement.

The authors would like to acknowledge support from KVI technical staff. O. Dermois and L. Huisman have contributed significantly for design and set up of the experiment. This research was supported by the Nederlandse Organisatie voor Wetenschappelijk Onderzoek (NWO) and the Stichting voor Fundamental Onderzoek der Materie (FOM) under program 114 ( \trimp) and FOM projectruimte 06PR2499.

\bibliography{isotope-shift-modified}

\end{document}